# Image-based retrieval of all-day cloud physical parameters for FY4A/AGRI and its application over the Tibetan Plateau

Zhijun Zhao[a, b], Feng Zhang[a, b, *], Wenwen Li[a], Jingwei Li[a, b]

[a]CMA-FDU Joint Laboratory of Marine Meteorology, Department of Atmospheric and Oceanic Sciences & Institutes of Atmospheric Sciences, Fudan University, Shanghai 200438, China.

[b]Key Laboratory for Information Science of Electromagnetic Waves, Ministry of Education, School of Information Science and Technology, Fudan University, Shanghai 200438, China.

Corresponding author: F. Zhang (fengzhang@fudan.edu.cn)

**Abstract**

Satellite remote sensing serves as a crucial means to acquire cloud physical parameters. However, existing official cloud products derived from the advanced geostationary radiation imager (AGRI) onboard the Fengyun-4A geostationary satellite suffer from limitations in computational precision and efficiency. In this study, an image-based transfer learning model (ITLM) was developed to realize all-day and high-precision retrieval of cloud physical parameters using AGRI thermal infrared measurements and auxiliary data. Combining the observation advantages of geostationary and polar-orbiting satellites, ITLM was pre-trained and transfer-trained with official cloud products from advanced Himawari imager (AHI) and Moderate Resolution Imaging Spectroradiometer (MODIS), respectively. Taking official MODIS products as the benchmarks, ITLM achieved an overall accuracy of 79.93% for identifying cloud phase and root mean squared errors of 1.85 km, 6.72 μm, and 12.79 for estimating cloud top height, cloud effective radius, and cloud optical thickness, outperforming the precision of official AGRI and AHI products. Compared to the pixel-based random forest model, ITLM utilized the spatial information of clouds to significantly improve the retrieval performance and achieve more than a 6-fold increase in speed for a single full-disk retrieval. Moreover, the AGRI ITLM products with spatiotemporal continuity and high precision were used to accurately describe the spatial distribution characteristics of cloud fractions and cloud properties over the Tibetan Plateau (TP) during both daytime and nighttime, and for the first time provide insights into the diurnal variation of cloud cover and cloud properties for total clouds and deep convective clouds across different seasons.

## 1. Introduction

Clouds play a profound influence on the radiation budget and climate change dynamics in the Earth-atmosphere system (Sassen et al., 2007; Loeb et al., 2014; Zhao and Garrett, 2015), the accurate acquisition of cloud physical parameters is essential for estimating shortwave radiation and advancing climate change research (Wang and Iwabuchi, 2019; Liang et al., 2021; Tana et al., 2023).

With the development of remote sensing technology, satellite remote sensing has become a primary means to acquire cloud physical parameters. Cloud products used in the field of satellite remote sensing are primarily derived from polar-orbiting and geostationary satellite sensors. For instance, the Moderate Resolution Imaging Spectroradiometer (MODIS) onboard the Aqua and Terra polar-orbiting satellites has been observing global clouds for more than 20 years. It has a lower orbit altitude of approximately 700 km compared to the geostationary satellite sensors, allowing for providing more accurate remote sensing datasets. As a result, these datasets are widely used to analyze the impact of clouds on climate effects (McCoy et al., 2017; Saponaro et al., 2020; Zheng et al., 2021). However, MODIS can only scan a small granule area (about 2000 km × 2330 km) during each satellite transit, and cannot provide continuous cloud observation datasets over larger spatial coverage (King et al., 2013). On the other hand, the advanced Himawari imager (AHI) and the advanced geostationary radiation imager (AGRI), which are onboard the Himawari-8 (H8) and Fengyun-4A (FY4A) geostationary satellites, respectively, have continuous cloud observation capabilities over larger areas. Nevertheless, AHI official only provides cloud products in daytime, and AGRI official lacks important cloud micro-physical property products.

The limitations in retrieving cloud physical parameters from geostationary satellite measurements can mainly be attributed to deficiencies in traditional retrieval algorithms. Most of these algorithms are based on the visible and shortwave infrared bi-spectral channels, which are effective in retrieving cloud properties during the daytime, but not applicable in sun glint regions (Nakajima et al., 1990; Letu et al., 2020; Zhou et al., 2022; Tana et al., 2023). While the traditional optimal estimation method based on infrared multi-channel measurements and infrared split-window method can be used for cloud retrieval at night, they have limitations in accurately estimating the properties of optically thick clouds due to the limited penetration ability of thermal infrared radiation (Inoue, 1985; Iwabuchi, 2014; Wang et al., 2016; Iwabuchi et al., 2018).

With advancements in artificial intelligence technology, machine learning models have been extensively applied in the development of cloud retrieval algorithms, significantly improving the efficiency and accuracy of estimating cloud properties during both daytime and nighttime (Minnis et al., 2016; Min et al., 2020; Yang et al., 2022; Lin et al., 2022; Li et al., 2022; Wang X. et al, 2022). Notably, the image-by-image (image-based) deep learning models have demonstrated remarkable improvements in estimating the properties of optically thick clouds by leveraging spatial information of clouds (Wang Q. et al, 2022; Zhao et al., 2023). However, these image-based models are currently limited to the application of satellites' sensors with similar orbit altitudes, thereby overlooking the untapped potential of combining the observation advantages of stationary and polar-orbiting satellites for cloud retrieval. Therefore, a transfer-learning-based approach has been proposed to achieve all-day retrieval of cloud properties from AHI thermal infrared measurements, yielding a set of cloud products with precision comparable to MODIS data (Tong et al, 2023; Li et al., 2023).

The Tibetan Plateau (TP), known as the highest plateau globally, has a unique plateau climate that significantly influences extreme weather formation in surrounding regions and global climate patterns (Liu et al., 2006; Zhou et al., 2021). Comprehending the changes in clouds over the TP holds great significance for numerical weather prediction and atmospheric circulation analysis (Yan et al., 2016). Although previous studies have investigated cloud changes over the TP using existing official satellite cloud products and reanalysis data (Lei et al., 2020; Ma et al., 2021; Wang et al., 2021), there is still a lack of analyses based on high-precision and high-frequency cloud products derived from stationary satellite measurements that cover the entire TP. Specifically, the coverage of AHI is restricted to the eastern region of the TP beyond 80°E, and AHI official cloud products are only available between BJT_11:00 and 17:00 during the daytime. The MODIS offers cloud observation data over a granule region of the TP during the transit time of the Aqua and Terra satellites, but it can only provide cloud optical properties during the daytime. Only AGRI has the advantage of comprehensiveness in TP coverage, it is limited to providing official cloud physical products with high precision and efficiency.

To overcome these limitations, an image-based transfer learning model (ITLM) was developed to achieve all-day retrieval of cloud phase (CLP), cloud top height (CTH), cloud effective radius (CER), and cloud optical thickness (COT) for FY4A/AGRI by effectively combining the observation advantages of polar-orbiting and geostationary satellites. The ITLM not only captures the continuous cloud changes observed by AHI but also improves the precision of retrieved cloud products to approach the level of official MODIS cloud products. Using the cloud products with spatiotemporal continuous and high precision retrieved from the ITLM, we conducted a comprehensive analysis of cloud cover and cloud properties over the TP, focusing on their spatial distribution and diurnal cycle.

The remaining contents of this study are structured as follows: Data sources and preprocessing are introduced in Section 2; The transfer learning model, evaluation and analysis method, and image mosaic method are described in Section 3; Section 4 demonstrates the evaluation results of cloud products by the polar-orbiting satellite data; Section 5 summarizes the key conclusions.

## 2. Data sources

### 2.1 Geostationary satellite data

The geostationary meteorological satellites H8 and FY4A were launched above the equator at 104.7°E and 140.7°E in October 2014 and December 2016, respectively. H8/AHI covers the range of 80°E to 200°E and 60°S to 60°N, while FY4A/AGRI covers the range of 24.12°E to 185.28°E and 80.57°S to 80.57°N. These satellites can observe the Earth every 10 minutes and 15 minutes using 14 and 16 channels, respectively (Bessho et al., 2016; Yang et al., 2017; Chen et al., 2019; Wang et al., 2022). In this study, the brightness temperatures (BTs) from bands 9 to 14 (central wavelength at 6.25, 7.1, 8.5, 10.8, 12.0, and 13.5 μm) and satellite zenith angle (SZA) of AGRI were used as input images for the model. Moreover, AHI official cloud products during the daytime were employed as the pre-trained target images for the model. The CLP and CTH from AGRI official products were used for evaluation and comparison with other cloud products.

### 2.2 Polar-orbiting satellite data

MODIS is carried on the polar-orbiting satellites Terra and Aqua, which were launched by the United States in December 1999 and April 2002, respectively. It uses 36 medium-resolution spectral bands (wavelengths ranging from 0.4 to 14.4 µm) to capture observation images of global land and ocean temperature, clouds, aerosol, water vapour and fire every 1-2 days (Barnes et al., 2003; King et al., 2013). Cloud Profiling Radar and Cloud-Aerosol Lidar with Orthogonal Polarization (CALIOP), onboard the Cloud-Aerosol Lidar and Infrared Pathfinder Satellite Observation (CALIPSO) satellite, combines active lidar instrumentation with passive infrared and visible imagers to detect the global vertical structure and characteristics of thin clouds and aerosols (Winker, et al.,2009; Saito et al., 2017). For the transfer-training of the model, MODIS official cloud products were used as the target images. When evaluating precision of other cloud products, the MODIS and CALIOP official cloud products were also used as the benchmarks.

## 2.3 Auxiliary data

Considering the generation mechanism and development process of clouds, we incorporated auxiliary data such as meteorological fields and surface emissivity (SE) as additional input images for the model. In this study, the meteorological fields data is derived from the fifth generation of atmospheric reanalysis produced by the European Centre for Medium-Range Weather Forecasts (ERA5), which provides hourly air temperature profiles (ATP), relative humidity profiles (RHP), surface skin temperature (SKT), and total column water vapour (TCWV) data with a spatial resolution of 0.25° x 0.25° (Hersbach et al., 2020; Bell et al., 2021). To prevent the redundancy of input images from reducing the the training efficiency of the model, AHP and RHP data were selected only at 4 pressure levels of 1000 hPa, 850 hPa, 500 hPa and 300 hPa were selected. Furthermore, SE data is derived from the Global Climate Modeling Grid Products of MODIS with a temporal resolution of 8 days and a spatial resolution of 0.05° x 0.05°.

## 3. Methodology

## 3.1 Data preprocessing

Table 1 provides an overview of the data used for training and evaluation in this study. The input and target images were obtained from different sources and had varying spatiotemporal resolutions. To ensure consistency, we performed data unification by aligning datasets onto a standardized latitude and longitude grid before feeding them into the model. We used the nearest interpolation method to resample AGRI BTs and MODIS official cloud products data, and employed the bilinear interpolation method to increase the spatial resolution of the ERA5 meteorological fields. Considering the differences in observation range and frequency among different satellite sensors, we selected hourly images within the overlapping region (80° E-182.4 °E and 42.4°S-60°N) that covers both AGRI and AHI observation ranges. However, the image size of this overlapping region was too large to directly employ as input for the model. Consequently, we divided each full-disk image into multiple small sample images with a size of 64×64 for model training and testing. During the model application, we can reconstruct the full-disk image using an image mosaic method, as described in the supplementary materials.

Table 1. Summary of used data in this study

| | Variable | Source | Spatial Resolution | Temporal Resolution |
|---|---|---|---|---|
| Input | Brightness Temperature (from 6 bands at 6.25, 7.1, 8.5, 10.8, 12.0, and 13.5 μm) Satellite Zenith Angle | FY4A/AGRI | 4 km | 15 min |
| | Air Temperature Profile (from 4 pressure levels at 1000hPa，850hPa，500hPa and 300hPa) Relative Humidity Profile (from 4 pressure levels at 1000hPa，850hPa，500hPa and 300hPa) Surface Temperature (at 2 m) Total Column Water Vapor | ERA5 | 0.25° | 1h |
| | Surface Emissivity (from 6 bands at 3.75 μm, 3.96 μm, 4.05 μm, 8.55 μm, 11.03 μm and 12.02 μm) | Aqua/MODIS | 0.05° | 8 days |
| Target | Cloud Phase Cloud Top Height | H8/AHI (Pre-training) | 0.05° | 10 min |

| Evaluation | Cloud Effective Radius | Aqua and Terra/MODIS | 1km | 5 min |
| --- | --- | --- | --- | --- |
| | Cloud Optical Thickness | (Transfer-training) | | |
| | Cloud Phase | | | |
| | Cloud Top Height | Aqua and Terra/MODIS | 1 km | 5 min |
| | Cloud Effective Radius | | | |
| | Cloud Optical Thickness | | | |
| | Cloud Phase | CALIPSO/CALIOP | 1 km | --- |
| | Cloud Top Heigh | | | |

### 3.2 Image-based transfer learning model

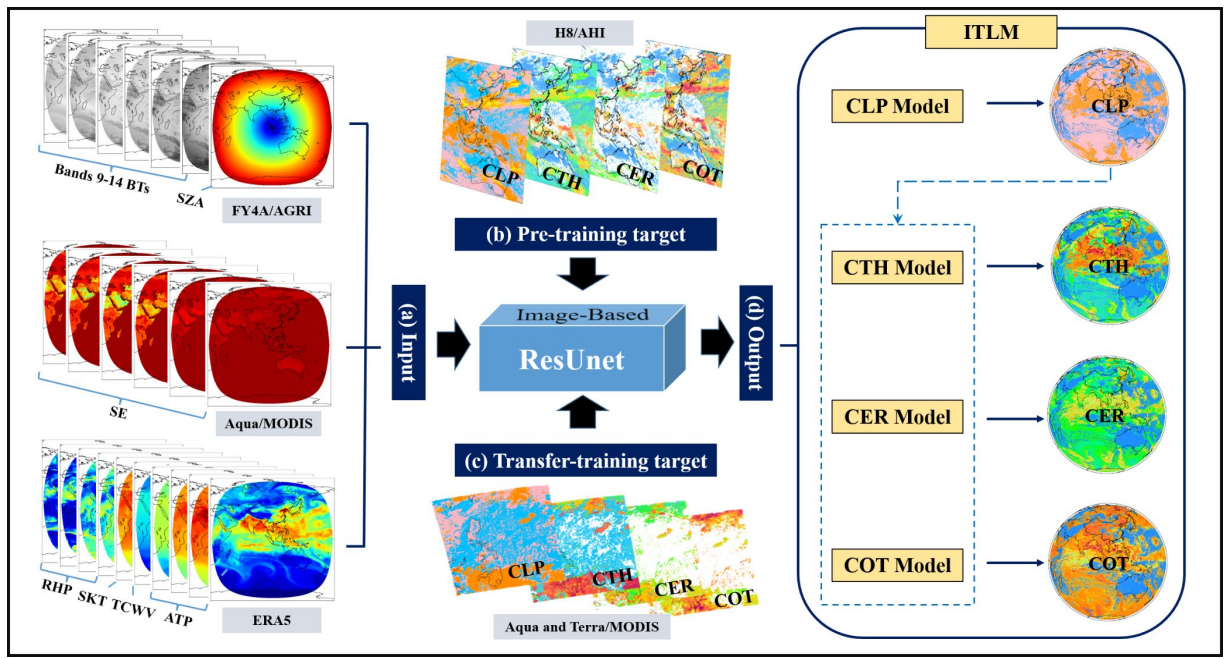

**Fig. 1.** The framework and flowchart of image-based transfer learning model (ITLM).

The ITLM is constructed using ResUnet, a deep learning network. It consists of two main stages: pre-training and transfer-training. In our previous study, we provided a comprehensive description of ResUnet's internal architecture, demonstrating its reliability and applicability for cloud retrieval (Zhao et al., 2023). In the pre-training stage, a preliminary relationship between input images and target images was established using ResUnet, which can obtain a pre-trained model using AHI official cloud products (Fig. 1a and b). In the transfer-training stage, we froze certain parameters of the pre-trained model and performed fine-tuning to establish the final correspondence between input images and MODIS target images (Fig. 1a and c). It should be noted that CLP products were used as new input images for training other models (Fig. 1d). As a result, the CLP model has input images for 23 channels, while other models have input images for 24 channels.

Table S1 information on the sample number and parameter setting used in the pre-training and transfer-training stages for different cloud parameter models. The training dataset consisted of sample images in 2019, while the testing dataset comprised sample images in 2020. Since the target images in the pre-training and transfer-training stages were obtained from different satellite sensors, the number of samples used for training may vary. To ensure consistency across the different cloud parameter models, we maintained identical parameter settings in both training stages. The optimal model parameters were determined through comprehensive comparison of various parameter selections, as described in our previous study (Zhao et al., 2023). For the four cloud parameter models, the learning rate and epoch were set to 0.001 and 400, respectively. Essentially, the training of the CLP model is a classification task with a larger number of samples, using the CrossEntropyLoss loss function and a batch size of 128. And the training of other cloud parameter models, which involve regression tasks with relatively smaller sample numbers, use the MSELoss loss function and a batch size of 64.

### 3.3 Evaluation and analysis method

To evaluate the effectiveness of the image-based model in extracting cloud spatial information and improving cloud retrieval performance, we conducted a comparative analysis between the ITLM and the pixel-based Random Forest model (PRFM) using an equally distributed testing dataset in 2020. Then, we used MODIS official products in 2020 as the benchmarks to assess the precision of CLP, CTH, CER, and COT obtained from AGRI ITLM, AGRI official and AHI official products across different spatiotemporal scales. CALIOP official products in 2020 were also utilized to evaluate the precision of CLP and CTH obtained from AGRI ITLM, AGRI official and AHI official products during the nighttime and various seasons. Additionally, we conducted an extensive analysis of cloud products obtained from different data sources over the TP, highlighting the spatiotemporal discontinuity of cloud products renders them unrepresentative. Based on the cloud products with spatiotemporal continuity and high precision retrieved from ITLM, we accurately analyzed the spatial distribution of cloud cover and cloud properties over the TP, and examined their diurnal variation characteristics across different seasons for the first time. Specifically, we selected January, April, July, and October in 2020 to represent winter, spring, summer, and autumn, respectively.

To evaluate the performance of CLP identification and cloud detection (CLD) by different products, we used overall accuracy (OA) and confusion matrix. Here, mixed or uncertain clouds were excluded from the evaluation, as these were not available in CALIOP and MODIS official products. Furthermore, we compared the probability density distribution to evaluate the precision of CTH, CER and COT by different products. A darker color near the diagonal represents denser pixel distributions, indicating higher precision in the cloud properties products. Additionally, we calculate correlation coefficient (R), mean absolute error (MAE), mean bias error (MBE), and RMSE to further quantify the precision of cloud products. Larger OA and R values, and smaller RMSE, MAE, and MBE values indicate higher precision.

## 4. Results and Discussion

### 4.1 Evaluation of image-based model against pixel-based model

To illustrate the advantages of the image-based model in cloud retrieval, we first compared the spatial distribution of cloud physical parameters obtained from AGRI official, AHI official, AGRI PRFM, and AGRI ITLM products. We randomly selected different numbers of pixels (200,000, 500,000, 1,000,000, and 1,500,000) from the ITLM training dataset to evaluate the performance of PRFM on a testing dataset of 200,000 random pixels. The results showed that a training dataset of 500,000 pixels was sufficient for PRFM training. Ultimately, the hyperparameters of PRFM were determined using grid search and cross-validation (GridSearchCV) method, with n_estimators = 180, max_depth = 40, min_samples_split = 3, and min_samples_leaf = 1.

Figure 2 illustrates the spatial distribution of cloud physical parameters at UTC_00:00 on April 26, 2020. Notably, the partial observation area was in the nocturnal condition at this time (Fig. 2b), and the sun glint angle (SGA) of AGRI and AHI was shown in Fig. 2i and Fig. 2m, respectively. We found that AGRI official products only provided the spatial distribution of all-day CLP and CTH (Fig. 2a and e), while AHI official products only provided the spatial distribution of cloud physical parameters during the daytime and had limitations in regions with SGA ≤ 30° (Fig. 2b, f, j and n). Both PRFM (Fig. 2c, g, k, and o) and ITLM (Fig. 2d, h, l, and p) were capable of retrieving cloud physical parameters at night and in regions with SGA ≤ 30°, and also providing cloud optical and micro-physical properties products for AGRI. There was significantly similar spatial distribution of cloud physical parameters between AGRI PRFM products and AGRI ITLM products, but the computational time of PRFM (~ 310 seconds) was more than 6 times longer than ITLM (~ 50 seconds) for a single full-disk retrieval image.

Furthermore, we compared the performance of ITLM and PRFM on an independent testing dataset in 2020. The OA of CLP identification between AGRI ITLM and MODIS official products reached 79.93%. However, there were still misjudgements with approximately 15% between clear sky and water clouds, as well as between water clouds and ice clouds. Nonetheless, ITLM demonstrated a commendable performance in identifying CLP, with accuracy exceeding 82% for clear sky and ice clouds (Fig. 3a). The joint probability density distribution of cloud properties between AGRI ITLM and MODIS official products showed close agreement along the diagonal. Despite the slight underestimation of ITLM presence (MBE<0), the errors of CTH, CER, and COT estimation were all within acceptable ranges. Specifically, the RMSE (MAE) were 1.85 km (1.26 km), 6.72 μm (5.85 μm), and 12.79 (8.05), respectively. Furthermore, the R also reached 0.884, 0.805, and 0.723, respectively, indicating the excellent performance of ITLM in cloud retrieval.

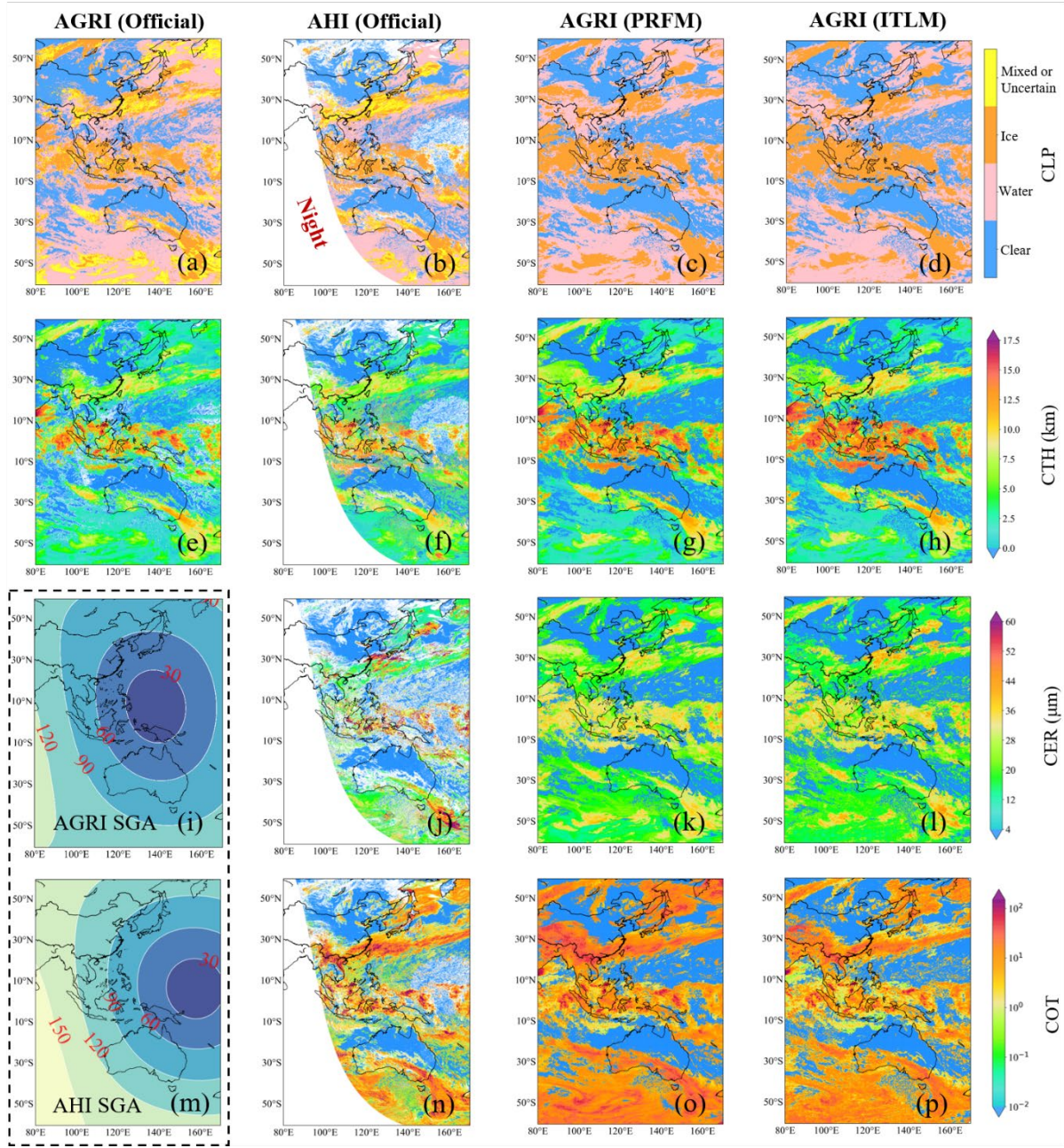

**Figure 2.** Spatial distribution of CLP (a-d), CTH (e-h), CER (j-l), and COT (n-p) obtained from AGRI official (a and e), AHI official (b, f, j and n), AGRI PRFM (c, g, k, o), and AGRI ITLM (d, h, l and p) products at UTC_00:00 on April 26, 2020. Sun glint angle (SGA) of AGRI (i) and AHI (m) is calculated by satellite zenith angle, satellite azimuth angle, solar zenith angle, and solar azimuth angle of FY4A and H8, respectively.

In comparison, the performance of PRFM in cloud retrieval was significantly degraded. The OA of CLP identification dropped to 75.73%, with lower accuracy for clear sky (77.21%), water clouds (72.85%), and ice clouds (78.30%). This clearly demonstrated that capability of PRFM in distinguishing different CLP was inferior to that of ITLM (Fig. 3b). The joint probability density distribution of cloud properties between AGRI PRFM and MODIS official products showed less concentration around the diagonal compared to AGRI ITLM products. The RMSE (MAE) of CTH, CER, and COT estimation increased to 2.33 km, 7.82 μm, and 18.81 (1.45 km, 6.17 μm, and 11.02), and the R decreased to 0.871, 0.740, and 0.690, respectively. These results indicated that PRFM had inferior performance in retrieving cloud properties compared to ITLM (Fig. 3 d, f and h).

The different performance of ITLM and PRFM in cloud retrieval can be attributed to their different feature extraction and target learning. ITLM, based on a convolutional neural network framework, captures extensive spatial information of clouds by establishing an image-by-image mapping relationship between input images and target images. Conversely, PRFM establishes a pixel-by-pixel mapping relationship between input features and target features, resulting in a loss of spatial information for the clouds. Clouds exist as a continuous entity in the atmosphere, and their spatial structure information can effectively improve the application effect of artificial intelligence technology on cloud retrieval.

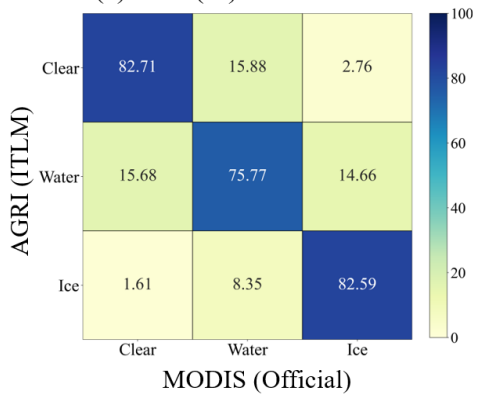
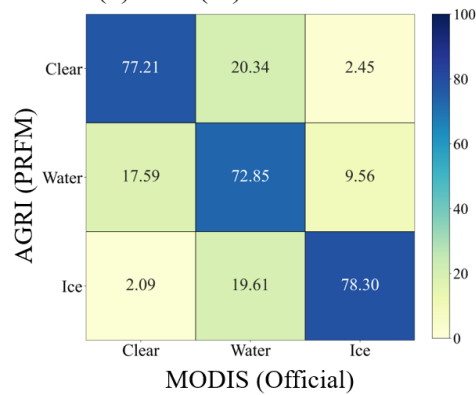
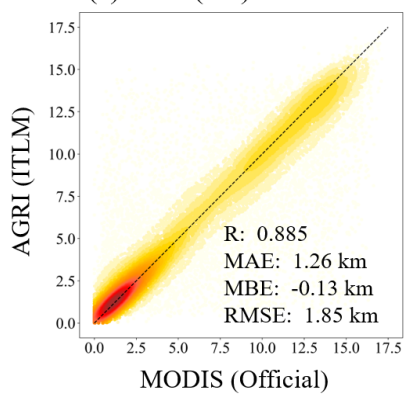
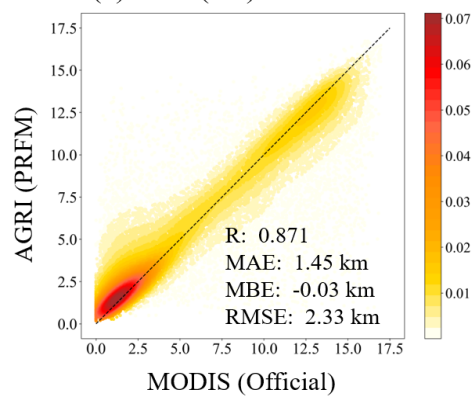
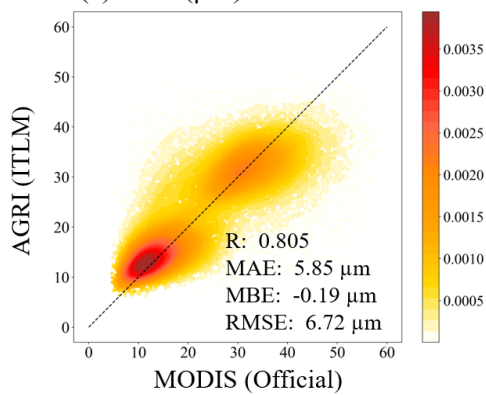
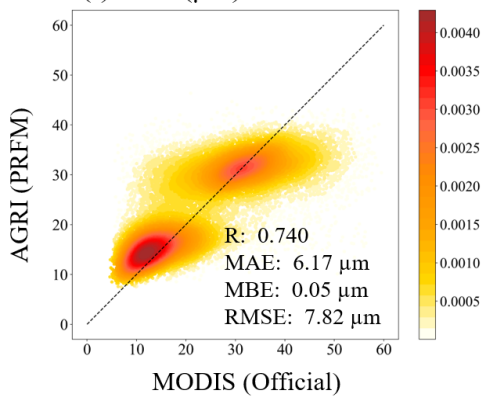
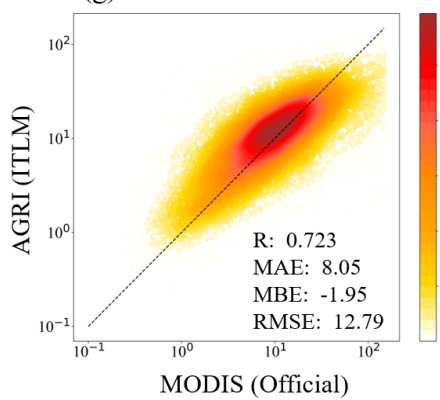
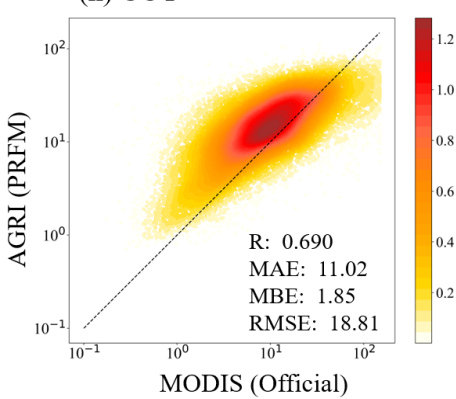

**Figure 3.** Confusion matrix of CLP (a, b) and joint probability density distribution of CTH (c, d), CER (e, f), and COT (g, h) between MODIS official and AGRI ITLM (a, c, e, g) or AGRI PRFM (b, d, f, h) products.

## 4.2 Evaluation of cloud products against MODIS data

To further illustrate the advantages of ITLM after transfer-training stage, we evaluated the precision of cloud products from various perspectives. In our previous study, we compared the precision of CLP, CTH, CER and COT obtained from AGRI official, AHI official and AGRI pre-trained model products with MODIS official data as the benchmarks (Zhao et al., 2023). Here, we compared previous results with the precision of AGRI ITLM products, as shown in Table 2. We found that the RMSE and OA of CTH, CER, COT, and CLP from AGRI pre-trained model products were 2.79km, 11.30μm and 16.12 and 74.44%, respectively, which comparable to the RMSE and OA of AHI official (2.72 km, 10.14 μm, 14.62, and 77.36%). However, the precision of CLP and CTH from pre-trained model products was significantly higher than that of AGRI official products (OA = 71.77%; RMSE = 3.58 km). Compared to AGRI pre-trained model products, the RMSE (OA) of CTH, CER, and COT (CLP) from AGRI ITLM products decreased (increased) to 1.85km, 6.72μm and 12.79 (79.93%), respectively. These results showed that ITLM greatly improved the performance of deep learning models in cloud retrieval after transfer training. Importantly, the precision of AGRI ITLM products was significantly better than that of AHI official and AGRI official products.

**Table 2.** Evaluation of cloud products using MODIS data in 2020 as the benchmark.

| Dataset | CLP (OA) | CTH (RMSE) | CER (RMSE) | COT (RMSE) |
|---|---|---|---|---|
| AGRI (Official) | 71.77% | 3.58 km | --- | --- |
| AHI(Official) | 77.36% | 2.72 km | 10.14 μm | 14.62 |
| AGRI (Pre-trained Model) | 74.44% | 2.79 km | 11.30 μm | 16.12 |
| AGRI (ITLM) | 79.93% | 1.85 km | 6.72 μm | 12.79 |

To provide a more intuitive representation of the impact of ITLM on cloud retrieval after transfer training, we compared the spatial distribution of total cloud fraction (TCF) and average CTH, CER and COT from AGRI ITLM, MODIS official (transfer-training targets) and AHI official (pre-training targets) products from January, April, July and October in 2020, as shown in Fig. 4. The spatial distribution of TCF and average cloud properties from AGRI ITLM, MODIS official and AHI official products showed high similarities. From the spatial distribution of TCF, the frequency of cloud occurrence in northern China and Australia was low, while the frequency of cloud occurrence in the equator and high latitude regions was high. Regarding the spatial distribution of cloud properties, the average CTH and CER (COT) were significantly higher (lower) in low latitudes compared to average values in high latitudes, especially near the equator. These differences can be attributed to the abundance of water vapor, relatively high temperatures and strong convection, which promote intense cloud formation and development in low latitudes. The above analysis showed that the models successfully captured the spatial distribution characteristics of AHI official and MODIS official products during both the pre-training and transfer training stages. However, the correspondence relationship between MODIS official and AGRI ITLM products was better than that between MODIS official and AHI official products. AHI official products exhibited significant overestimation (underestimation) for TCF, CER, and COT (CTH) compared to MODIS official products.

The differences in the spatial distribution of TCF and average CTH, CER and COT among AHI official, MODIS official, and AGRI ITLM products are shown in Fig. 5. Positive differences are represented by warm colors, and negative differences are represented by cool colors. It was evident that the average difference between AHI official and MODIS official products was significantly larger than the average difference between AGRI ITLM and MODIS official products, indicating that the precision of the AGRI ITLM products was closer to that of MODIS official products. Except for CTH, the comparisons of TCF, CER, and COT between AHI official and MODIS official products were dominated by positive biases, suggesting that AHI official products tended to overestimate TCF, CER, and COT compared to MODIS official products (Fig. 5a, g and j). Regarding the average CTH, AHI official products displayed significant underestimation in high-cloud regions over China, Australia, and the equator, while showing significant overestimation in other low-cloud regions (Fig. 5d). Furthermore, the differences in spatial distribution between AHI official and AGRI ITLM products were similar to those between AHI official and MODIS official products, further highlighting that the close precision of AGRI ITLM products to the true targets during the transfer-training stage.

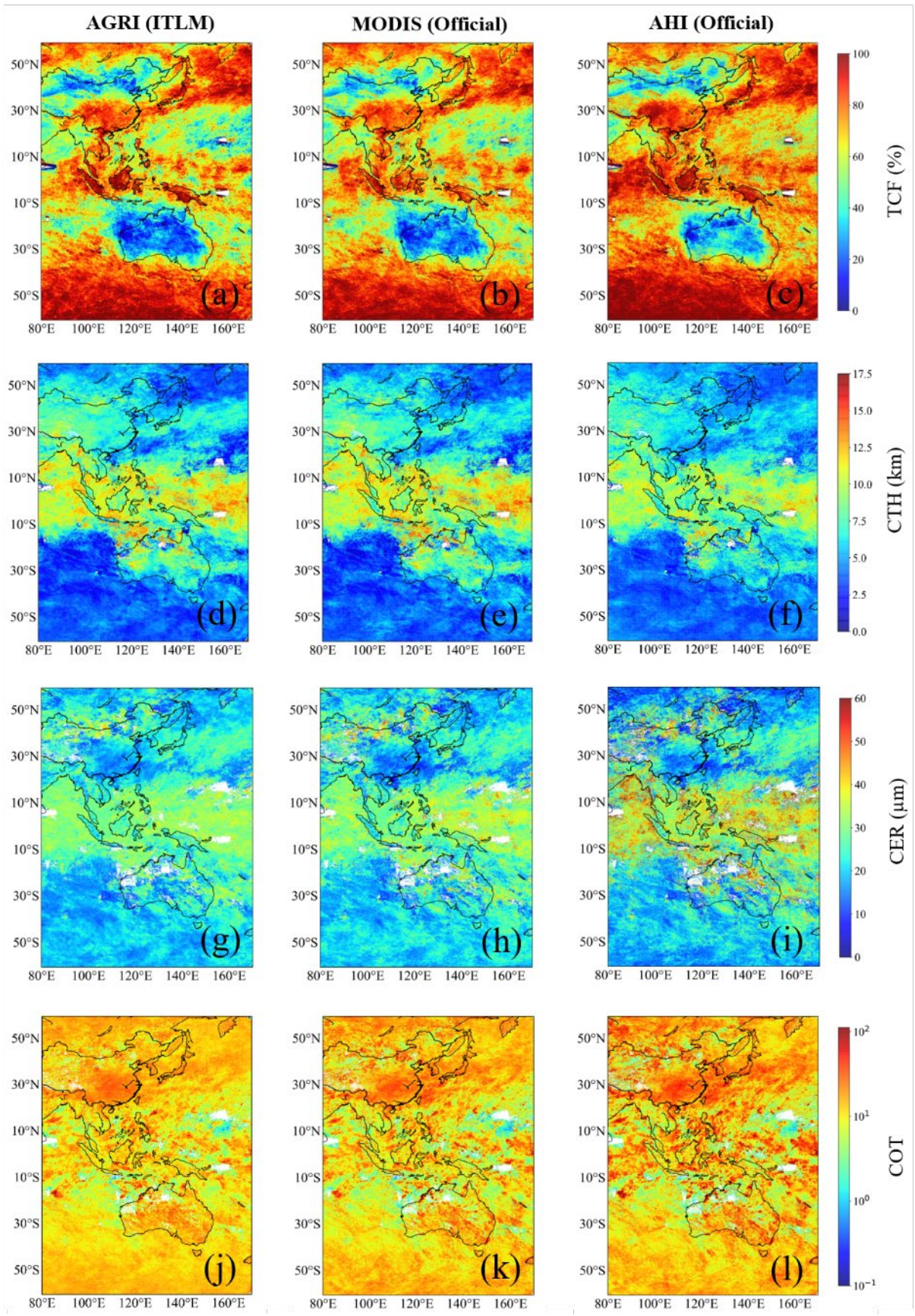

**Figure 4.** Spatial distribution of TCF (a-c) and average CTH (d-f), CER(g-l) and COT (j-l) obtained from AGRI ITLM (a, d, g, j), MODIS Official (b, e, h, k) and AHI Official (c, f, i, l) products in January, April, July, and October 2020.

Through the evaluation of cloud products against MODIS data at different temporal and spatial scales, we have confirmed that the ITLM not only captures spatial distribution features of AHI official products, but also provides all-day CLP, CTH, CER, and COT products for AGRI with precision closer to that of MODIS official products. This fully demonstrates that ITLM effectively combines the observation advantages of stationary and polar-orbiting satellites.

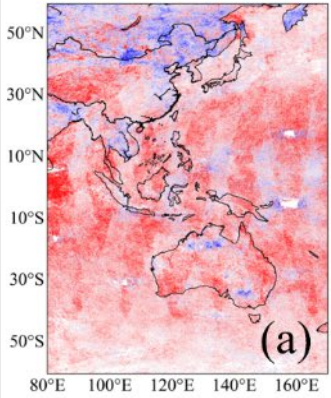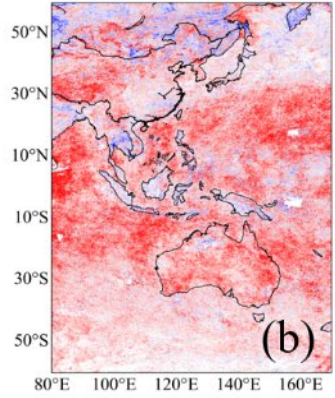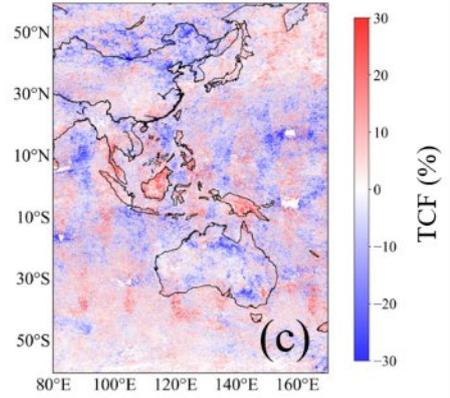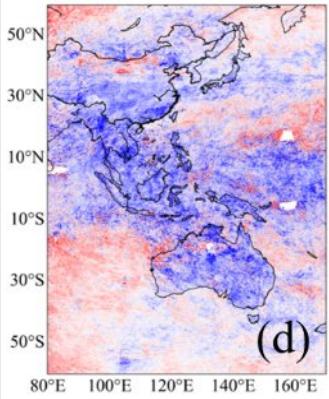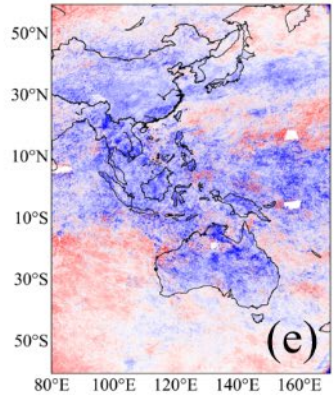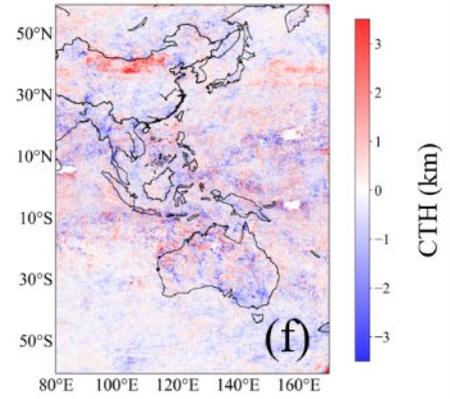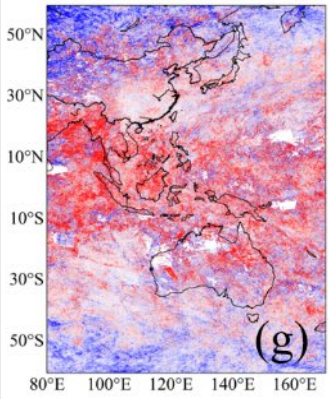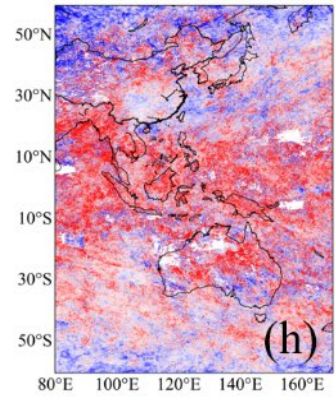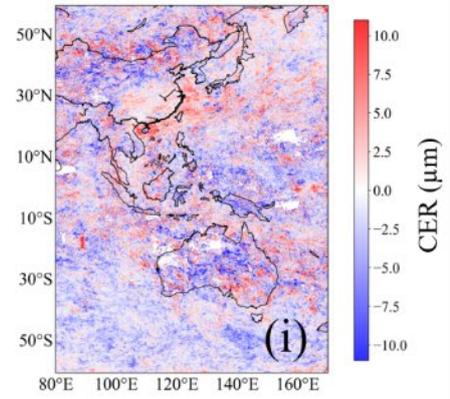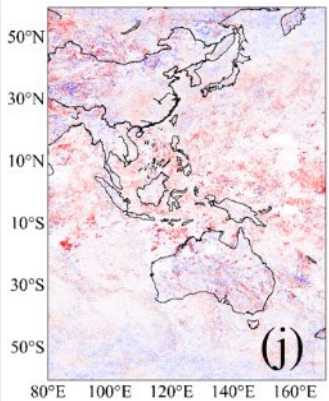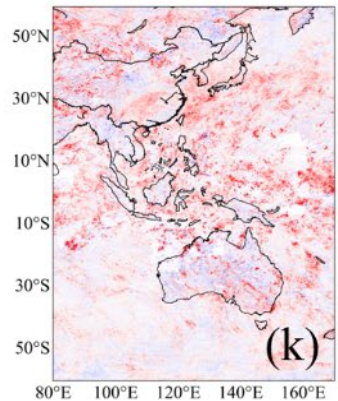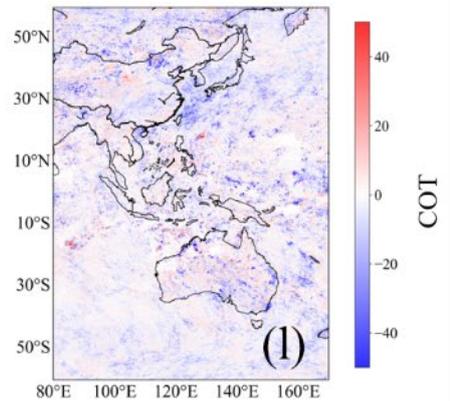

**Figure 5.** Differences in the spatial distribution of TCF (a-c) and average CTH (d-f), CER(g-l) and COT (j-l) between AHI official and MODIS official (a, d, g, j), AHI official and AGRI ITLM (b, e, h, k), AGRI ITLM and MODIS official (c, f, i, l) products in January, April, July, and October 2020.

### 4.3 Nighttime evaluation of cloud products against CALIOP data

To demonstrate the stability and reliability of ITLM, it is important to evaluate and compare the precision of the cloud products during the nighttime and daytime. Here, we utilized an active remote sensing dataset from CALIOP with a spatial resolution of 1 km, but it can only extract precise and direct cloud top information during the observation period of AGRI. Therefore, the evaluation focused on precision of CLP and CTH obtained from AGRI official, AHI official, and AGRI ITLM products during the nighttime and daytime in January, April, July, and October 2020.

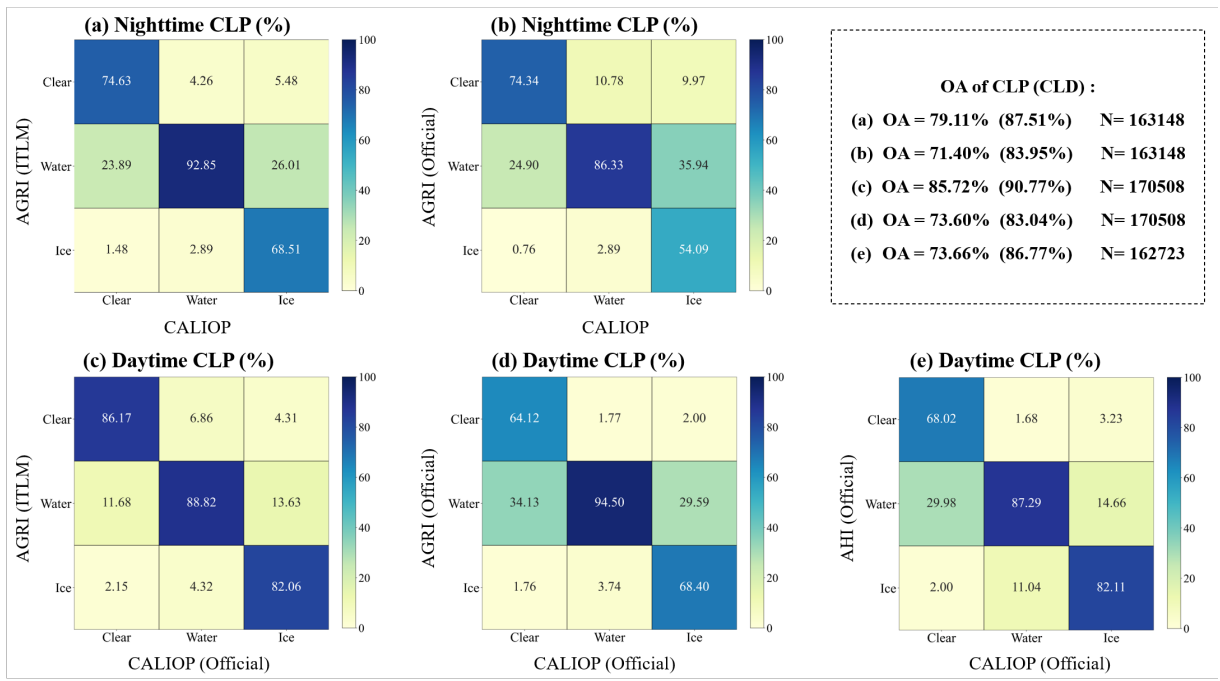

**Figure 6.** Confusion matrix of CLP between CALIOP official and AGRI ITLM (a and c), AGRI official (b and d), AHI official (e) during the nighttime (a and b) and daytime (c-e) in January, April, July, and October 2020.

Fig. 6 shows the confusion matrix of CLP obtained from AGRI official, AHI official and AGRI ITLM products against CALIOP data. The OAs of CLP identification by AGRI ITLM and AGRI official products at night were 79.11% and 71.40%, respectively (Fig. 6a-b). During the daytime, AGRI official and AHI official products had comparable OAs of approximately 73.60% for CLP identification, which was notably lower than the OA of 85.72% achieved by AGRI ITLM products (Fig. 6c-e). AGRI ITLM products consistently demonstrated higher precision than AGRI official and AHI official products during both nighttime and daytime, particularly for identifying water and ice clouds at night. In addition, the OAs of CLD by AGRI ITLM products were comparable during nighttime and daytime (87.51% and 90.77%), significantly surpassing those of AGRI official (83.04% and 83.95%) and AHI official (86.77%) products. These results illustrated the stability of ITLM in CLP identification and CLD at night.

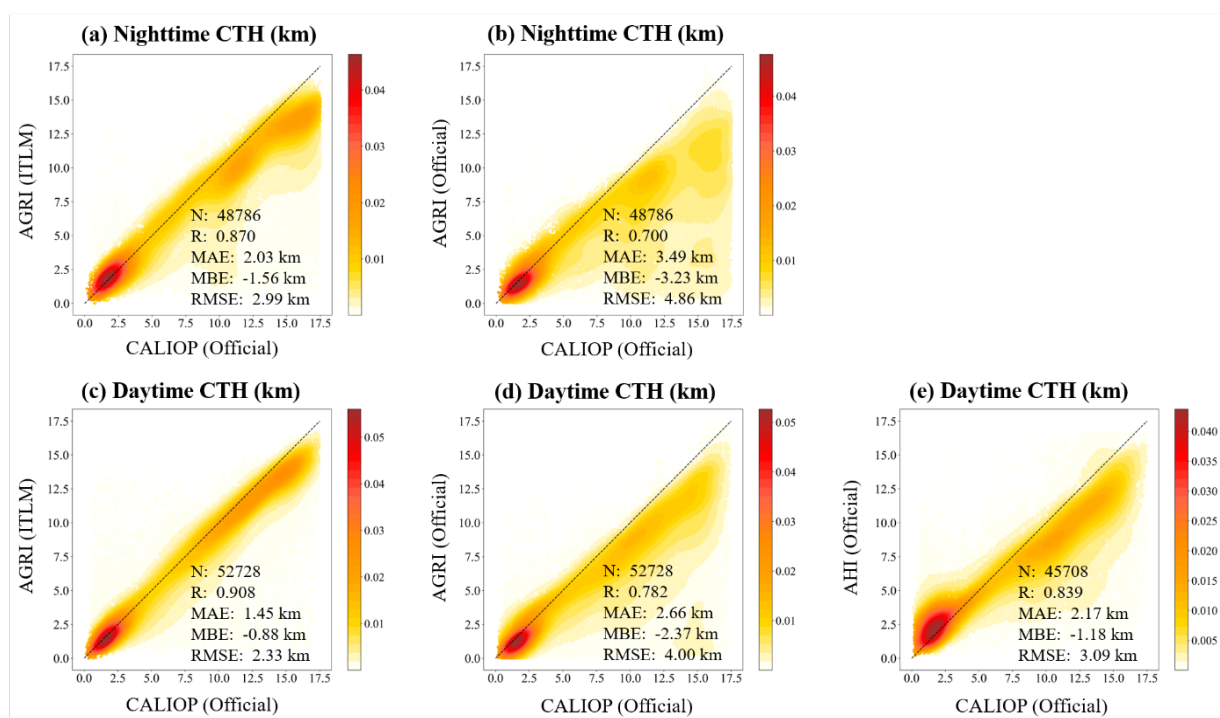

**Figure 7.** Joint probability density distribution of CTH between CALIOP official and AGRI ITLM (a and c), AGRI official (b and d), AHI official (e) during the nighttime (a and b) and daytime (c-e) in January, April, July, and October 2020. Color represents the occurrence frequency obtained by kernel density estimation, and the dashed black line indicates the diagonal line.

The joint probability density distribution of CTH obtained from AGRI official, AHI official and AGRI ITLM products against CALIOP data are shown in Fig.7. The distribution of CTH at night demonstrated that AGRI ITLM products exhibited a significantly higher concentration around the diagonal compared to AGRI official products (Fig. 7a and b). The RMSE and R of CTH from AGRI ITLM products were 2.99km and 0.870, respectively, which were significantly lower than the values of 4.86 km and 0.700 obtained from AGRI official products. Moreover, AGRI official products showed a significant underestimation of CTH (MBE: -3.23km), particularly for high clouds above 10 km, indicating that AGRI ITLM products provided more precise CTH information during the nighttime. The evaluation of CTH precision during the daytime confirmed the superior performance of AGRI ITLM products compared to AGRI official and AHI official products (Fig. 7c-e). Overall, AGRI ITLM products had better consistency with CALIOP data in CTH distribution during both daytime and nighttime.

Table 3 lists the evaluation results of CLP and CTH precision across spring, summer, autumn and winter in 2020 using CALIOP data as the benchmarks. The evaluation of CLP identification (CLD) by AGRI ITLM products showed consistent performance during both nighttime and daytime, with OAs ranging from 78.91% to 80.76% and 85.31% to 86.65% (86.25% to 88.25% and 89.78% to 91.53%), respectively. These results indicated that ITLM maintains stable performance in CLP identification over the seasonal cycle. In addition, the changes in RMSE and R of CTH from AGRI ITLM products were also not significant across different seasons, with RMSE ranging from 2.97 km to 3.08 km and 2.29 km to 2.44 km, and R ranges from 0.867 to 0.877 and 0.884 to 0.919 during nighttime and daytime, respectively. Compared to AGRI official and AHI official products, AGRI ITLM products exhibited higher CLP and CTH precision across all seasons, indicating the strong stability and reliability of ITLM throughout the seasonal cycle.

**Table 3.** Evaluation of CLP and CTH precision across different seasons using CALIOP data in 2020 as the benchmarks.

| Indicator | | Product | N/D | Spring | Summer | Autumn | Winter |
|---|---|---|---|---|---|---|---|
| CLP | OA (%) | AGRI (ITLM) | Nighttime | 78.91 | 79.83 | 80.02 | 80.76 |
| | | | Daytime | 85.53 | 85.31 | 86.65 | 85.35 |
| | | AGRI (Official) | Nighttime | 70.56 | 71.21 | 69.85 | 73.68 |
| | | | Daytime | 70.24 | 75.13 | 76.10 | 72.94 |
| | | AHI (Official) | Daytime | 72.47 | 74.74 | 74.32 | 73.10 |
| CLD | OA (%) | AGRI (ITLM) | Nighttime | 87.36 | 88.25 | 86.25 | 88.15 |
| | | | Daytime | 90.80 | 91.10 | 91.53 | 89.78 |
| | | AGRI (Official) | Nighttime | 82.84 | 85.97 | 83.08 | 83.97 |
| | | | Daytime | 81.62 | 85.02 | 84.59 | 81.29 |
| | | AHI (Official) | Daytime | 86.67 | 87.30 | 87.27 | 85.21 |
| CTH | RMSE (km) | AGRI (ITLM) | Nighttime | 2.97 | 2.99 | 3.08 | 2.91 |
| | | | Daytime | 2.44 | 2.29 | 2.38 | 2.33 |
| | | AGRI (Official) | Nighttime | 4.97 | 4.79 | 4.82 | 4.86 |
| | | | Daytime | 4.40 | 3.78 | 3.98 | 3.78 |
| | | AHI (Official) | Daytime | 3.08 | 3.02 | 3.06 | 3.19 |
| | R | AGRI (ITLM) | Nighttime | 0.867 | 0.877 | 0.873 | 0.877 |
| | | | Daytime | 0.884 | 0.919 | 0.913 | 0.911 |
| | | AGRI (Official) | Nighttime | 0.641 | 0.717 | 0.690 | 0.756 |
| | | | Daytime | 0.762 | 0.764 | 0.803 | 0.786 |
| | | AHI (Official) | Daytime | 0.833 | 0.828 | 0.849 | 0.843 |

**4.4 Spatiotemporal distribution characteristics of clouds over the Tibetan Plateau**

To illustrate the advantages of AGRI ITLM products with spatiotemporal continuity and high precision, we compared the spatial distribution characteristics of AGRI ITLM products with MODIS official and AHI official products over the TP. Notably, the analysis focused on the areas between 63°E-105°E and 20°N-45°N, as defined by the TPBoundary_rectangle provided in the Integration Dataset of Tibet Plateau boundary, with specific attention given to altitudes exceeding 2500 meters (Zhang, 2019).

Figure 8 shows the spatial distribution of TCF, water cloud fraction (WCF), and ice cloud fraction (ICF) in 2020 over the TP. During the daytime, the spatial distribution characteristics of TCF, WCF, and ICF between MODIS official and AGRI ITLM products were highly consistent during the transit time of MODIS (Fig. 8a-f). However, when considering the entire daytime period from BJT_08:00 to BJT_18:00, AGRI ITLM products showed significant differences, particularly in the central region of the TP (Fig. 8g-i). The statistical results of AHI official products covering BJT_11:00 to BJT_17:00 also differed from MODIS official and AGRI ITLM products (Fig. 8m and n). It should be noted that most ice clouds are classified as mixed clouds or uncertain phase in AHI official products, resulting in ICF significantly lower than 40% (Fig. 8o). Through the above comparative analysis, it can be found that the spatial distribution of cloud fractions calculated based on MODIS official and AHI official products with spatiotemporal discontinuity were not representative over the TP.

Therefore, we analyzed the spatial distribution characteristics of cloud fractions over the TP during the daytime and nighttime based on AGRI ITLM products. As shown in Fig. 8g-l, TCF presented a consistent spatial distribution pattern throughout the day and night, with TCF exceeding 60% above 2500m. However, the spatial distribution of WCF and ICF demonstrated contrasting characteristics between daytime and nighttime. Over the region with an altitude above 2500m, WCF during the daytime was significantly lower than that at night, while ICF displayed the opposite characteristics. These findings indicated that the importance of considering cloud fractions at night when analyzing the seasonal and diurnal variation of cloud cover frequency (CCF) over the TP.

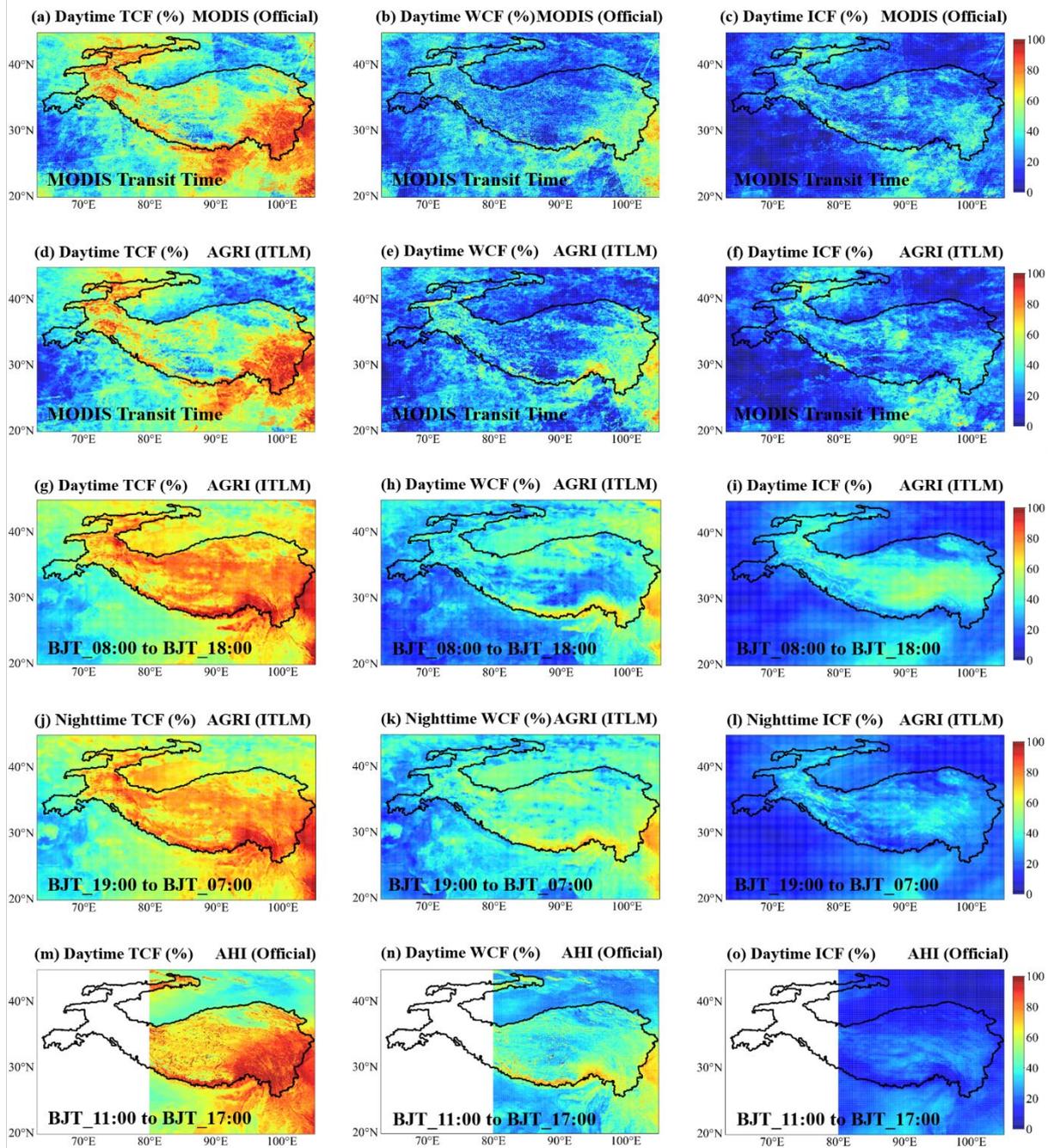

**Figure 8.** Spatial distribution of TCF (a, d, g, j and m), WCF (b, e, h, k and n), and ICF (c, f, i, l and o) during the daytime (a-i, and m-o) and nighttime (j-l) calculated by CLP obtained from MODIS (a-c) official, AGRI ITLM (d-l), and AHI official (m-o) products in January, April, July, and October 2020 over the TP. The black outline represents the area with an altitude ≥ 2500m.

Similarly, the spatial distribution characteristics of cloud properties in 2020 over the TP were analyzed. When using MODIS official products within the transit time, there was a high consistency in spatial distribution with AGRI ITLM products (Fig. 9a-f). However, when considering the entire daytime period from BJT_08:00 to BJT_18:00, AGRI ITLM products showed significantly lower CER and COT distribution compared to MODIS official products (Fig. 9g-i). Moreover, AHI official products, which had limited spatiotemporal coverage, exhibited completely different spatial distribution characteristics. These findings suggested that cloud products with spatiotemporal discontinuity cannot accurately represent the diurnal cycle of cloud properties over the TP.

To address this problem, a statistical analysis was conducted on the spatial distribution of average cloud properties based on AGRI ITLM products. During the daytime and nighttime, average CTH above 2500m was significantly higher than other surrounding regions. Due to the heating effect of solar radiation during the daytime, the maximum of average CTH reached 11km, which was significantly higher than that distributed at night (Fig. 9g and j). In addition, average CER was notably lower at night, ranging from 14 μm to 30 μm (Fig. 9h and k). In the spatial distribution of average COT, there was a similar distribution pattern between nighttime and daytime, with slightly higher values at night and a predominance of thick clouds with COT>10 (Fig. 9i and l). These findings illustrated the importance of considering cloud products during the nighttime for investigating the seasonal and diurnal cycles of cloud properties over the TP.

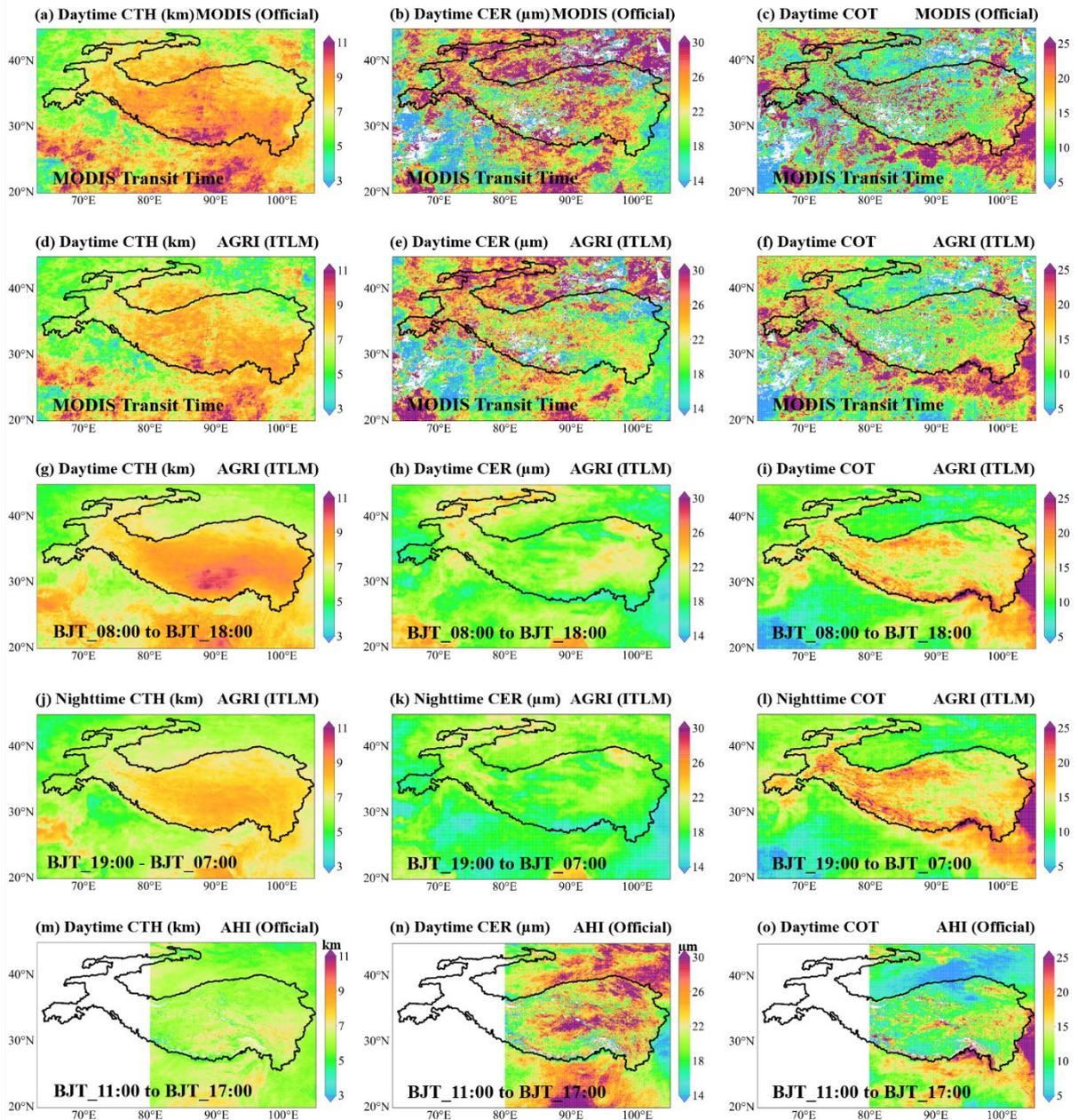

**Figure 9.** Spatial distribution of CTH (a, d, g, j and m), CER (b, e, h, k and n), and COT (c, f, i, l and o) during the daytime (a-i, and m-o) and nighttime (j-l) obtained from MODIS (a-c) official, AGRI ITLM (d-l), and AHI official (m-o) products in January, April, July, and October 2020 over the TP. The black outline represents the area with an altitude ≥ 2500m.

The relevant analysis of deep convective clouds over the TP is critical due to their close relationship with local severe weather events and their substantial impact on global weather and climate processes (Okamura et al., 2017; Li et al., 2021). Using the COT and CTH from AGRI ITLM products, we identified deep convectional clouds over the TP according to the cloud classification criteria defined by the International Satellite Cloud Climatology Project (Rossow and Schifer, 1999). Here, it was necessary to convert CTH into cloud top pressure. Then, we obtained the diurnal variations of average cloud cover frequency (CCF), CTH, CER and COT for total clouds and deep convective clouds across different seasons.

Overall, the analysis revealed distinct characteristics in the diurnal cycles of total clouds and deep convective clouds across different seasons (Fig. 10). The CCF of total clouds varied between 30% and 80% over the TP, with the maximum and minimum values occurring around BJT_08:00 and BJT_13:00, respectively. In the seasonal distribution, CCF showed consistent changes in summer and winter, with significantly higher values compared to spring and autumn (Fig.10a). The average CTH was smallest during the winter and remained relatively stable, while it was considerably higher during the summer, reaching an average of approximately 9 km. As convection system develops, CTH generally increased in the early afternoon, peaking around BJT_19:00. In the variation of the optical properties of total clouds, CER and COT displayed different characteristics. Specifically, CER exhibited an increasing trend after BJT_09:00, reaching its peak around BJT_17:00, while COT exhibited an opposite trend. Moreover, the average CER reached the maximum (~23 μm) in spring and the minimum (~15 μm) in autumn, with minor changes in winter. The average COT was smallest in spring (~11) and largest in summer (~14), with significant variations between autumn and winter, ranging from 10 to 17 (Fig. 10e and g).

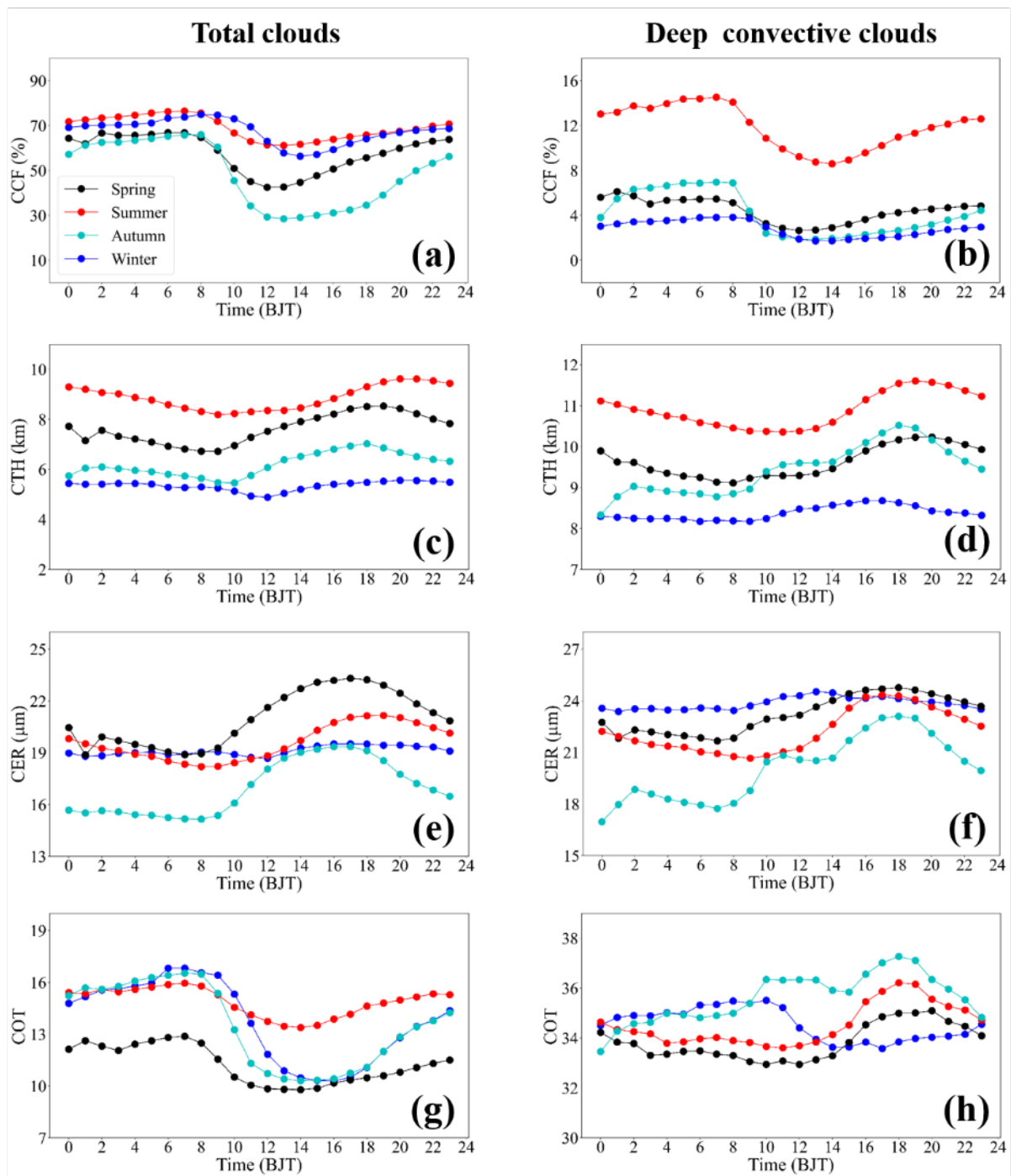

**Figure 10.** Diurnal variation curves of average CCF (a and b), CTH (b and f), CER (c and g) and COT (d and h) for total clouds (a, c, e and g) and deep convective clouds (b, d, f, h) across different seasons in 2020 based on AGRI ITLM products over the TP. Black, red, cyan, and blue lines represent spring, summer, autumn, and winter, respectively.

For the variation of deep convective clouds, CCF also peaked at BJT_ 08:00 and BJT_ 13:00 in the diurnal cycle, but showed significant differences in seasonal distribution compared to total clouds. In summer, the average CCF of deep convective clouds was highest among all seasons, exceeding 8% at each moment. Winter exhibited the smallest variability in CCF, with fluctuations of about 3% (Fig. 10b). The diurnal variation of CTH for deep convective clouds followed a similar pattern to total clouds, with an increase following enhanced convection in the early afternoon (Fig.10d). Seasonally, average CTH peaked in summer (~11 km) and reached its lowest point in winter (~8.5 km). In addition, the optical properties of deep convective clouds, specifically CER and COT, displayed varying diurnal variations across different seasons. In spring and summer, both CER and COT had an increased trend in the early afternoon, peaking around BJT_18:00. The average CER in summer was smaller than in spring, while the average COT showed the inverse situation. In autumn, there was unique characteristic with the minimum average CER (~20 μm) and maximum average COT (~35) among the seasons. During the winter, CER underwent minor changes throughout the diurnal cycle, while COT dropped significantly between BJT_10:00 and 14:00 (Fig. 10f and h).

According to the above analysis, it is evident that the spatial distribution characteristics of cloud fractions and cloud properties over the TP obtained from AHI official and MODIS official products with spatiotemporal discontinuity do not provide a representative depiction. Therefore, we accurately analyzed the spatial distribution characteristics of cloud fraction and cloud properties during the daytime and nighttime using AGRI ITLM products with spatiotemporal continuity and high precision. Additionally, we examined the diurnal variation of CCF and cloud properties across different seasons for the first time.

## 5. Conclusion

In this study, the ITLM was developed by using thermal infrared BTs from FY4A/AGRI and auxiliary data as predictors to realize the all-day retrieval of cloud physical parameters. This model combined the advantages of high-frequency observations from geostationary satellites and high-precision observations from polar-orbiting satellites, addressing the limitations of spatiotemporal discontinuities and low precision in existing official cloud products of FY4A/AGRI. Based on AGRI ITLM products, we investigated the spatial and temporal distribution characteristics of cloud fractions and cloud properties over the TP. The key conclusions are summarized as follows:

1. The ITLM outperforms the PRFM in terms of precision and efficiency in cloud retrieval. The ITLM utilizes the spatial information of clouds through image-by-image mapping between input images and target images, improving the retrieval performance. Additionally, the efficiency cost of the PRFM for a single full-disk retrieval is more than six times higher than that of the ITLM.
2. Compared to MODIS official products, AGRI ITLM products demonstrate significantly higher precision than AGRI official and AHI official products. Furthermore, ITLM shows stable performance during both nighttime and daytime across different seasons when compared to active remote sensing dataset from CALIOP.
3. The spatiotemporal distribution characteristics of cloud fractions and cloud properties over the TP obtained from MODIS official and AHI official products lack representation due to insufficient temporal and spatial coverage. Therefore, based on AGRI ITLM products with spatiotemporal continuity and high precision, the spatiotemporal variations of clouds over the TP are comprehensively and accurately analyzed. Importantly, diurnal variation characteristics of cloud cover and cloud properties for total clouds and deep convective clouds are presented for the first time.

Future research should focus on further analyzing and explaining the spatiotemporal distribution characteristics of clouds over the TP by investigating the interactions between clouds, atmospheric circulation, and unique topographic features. Furthermore, we plan to generate a all-day, high-precision, and high-frequency dataset of cloud physical parameters from March 2018 to March 2023 for FY4A geostationary satellite, which will be made publicly available for applications and research in East Asia, China, the Pacific Program and other relevant regions. Although ITLM has demonstrated excellent performance in cloud retrieval, the precision of this new algorithm still limited by the quality of MODIS official products. To overcome this limitation, deep learning models need to be further explored and developed by utilizing the active remote sensing dataset as the target images and incorporating the time evolution information of clouds. Of course, the use of advanced artificial intelligence techniques such as reinforcement learning and diffusion models can be considered to enhance the performance of deep learning models in cloud retrieval (Wang Q et al., 2022).

## Acknowledgements

This work has been supported by the National Natural Science Foundation of China (Grant 42075125 and Grant 42222506). We sincerely acknowledge the Fengyun Satellite Data Center and National Aeronautics, JAXA Himawari Monitor and National Aeronautics and Space Administration for providing AGRI official (http://data.nsmc.org.cn/PortalSite/Data/Satellite.aspx), AHI official (https://www.eorc.jaxa.jp/ptree/index.html), CALIOP official (https://subset.larc.nasa.gov/calipso/login.php), and MODIS official (https://ladsweb.modaps.eosdis.nasa.gov) products. In addition, we also acknowledge the European Centre for Medium-Range Weather Forecasts to provide ERA5 atmospheric reanalysis data (https://cds.climate.copernicus.eu). The datasets of Integration dataset of Tibet Plateau boundary is provided by National Tibetan Plateau/Third Pole Environment Data Center (http://data.tpdc.ac.cn).